\documentclass[aip,apl,reprint,amsmath,superscriptaddress]{revtex4-1}
\input{epsf.sty}
\usepackage{epsfig}
\usepackage{graphicx}
\usepackage{epstopdf} 
\usepackage{amsmath,amsthm,amssymb,latexsym,amsfonts,times,mathrsfs,verbatim, lipsum}
\usepackage{bm}
\usepackage{color}

\begin{document}
\title{Nanoscale Torsional Optomechanics} 

\author{P.H. Kim}
\author{C. Doolin}
\author{B.D. Hauer}
\author{A.J.R. MacDonald}
\affiliation{Department of Physics, University of Alberta, Edmonton, Alberta, Canada T6G 2E9}
\author{M.R. Freeman}
\affiliation{Department of Physics, University of Alberta, Edmonton, Alberta, Canada T6G 2E9}
\affiliation{National Institute for Nanotechnology, Edmonton, Alberta, Canada T6G 2M9}
\author{P.E. Barclay}
\affiliation{National Institute for Nanotechnology, Edmonton, Alberta, Canada T6G 2M9}
\affiliation{Institute for Quantum Information Science and Department of Physics and Astronomy,
University of Calgary, Calgary, Alberta, Canada T2N 1N4}
\author{J.P. Davis}\email{jdavis@ualberta.ca}
\affiliation{Department of Physics, University of Alberta, Edmonton, Alberta, Canada T6G 2E9}

\date{Version \today}

\begin{abstract}  
Optomechanical transduction is demonstrated for nanoscale torsional resonators evanescently coupled to optical microdisk whispering gallery mode resonators.  The on-chip, integrated devices are measured using a fully fiber-based system, including a tapered and dimpled optical fiber probe.  With a thermomechanically calibrated optomechanical noise floor down to 7 fm/$\sqrt{\textrm{Hz}}$, these devices open the door for a wide range of physical measurements involving extremely small torques, as little as $4\times10^{-20}$ N$\cdot$m.

\end{abstract}

\maketitle

The simplicity of the angular form of Hooke's Law, $\tau = \kappa\theta$, belies its outstanding versatility and usefulness.  Mechanical devices with small torsion constants ($\kappa$) have been designed to respond to torques associated with, and therefore sensitively measure, gravity, \cite{Cav98,Lut82} charge, \cite{Cle98} magnetism, \cite{Cha05,Dav10,Dav10b} the Casimir Force, \cite{Cha01} shear modulus,\cite{Kle85} superfluids, \cite{Bis80} the de Hass-van Alphen effect, \cite{Con64,Lup99} superconductors, \cite{Li07} persistent currents, \cite{Ble09} the angular momentum of light, \cite{Bet36} and more.  In most cases such measurements can be improved by scaling down the size of the torsional devices to reduce the torsion constant.  Often this scale-down presents opportunities for new physics associated with these smaller length-scales, such as mesoscopic phenomena. \cite{Dav10b,Ble09}  Hence there is a natural progression towards construction of nanoscale torsional resonators. \cite{Dav10}  But nanomechanical resonators suffer from their own set of difficulties, namely in sensitive transduction of their mechanical displacements using techniques that do not scale down well.\cite{Eki05}  In this manuscript, we present the first nanoscale torsional resonators to take advantage of optomechanical detection, \cite{Ane09,Eic09} which has broken through this limitation.\cite{Ane10} We demonstrate nanoscale torsional devices with a displacement noise floor down to 7 fm/$\sqrt{\textrm{Hz}}$, corresponding to an angular displacement of 4 nrad/$\sqrt{\textrm{Hz}}$ and a torque sensitivity of $4\times10^{-20}$ N$\cdot$m / $\sqrt{\textrm{Hz}}$.  For comparison, this torque sensitivity is just barely surpassed by optical \cite{Bry03} and magnetic \cite{Lip10} tweezers ($\sim$$10^{-21}$ N$\cdot$m) that are capable of single-molecule torque measurements.  In a modulated magnetic field ($\boldsymbol{H_{AC}}$) of 1 kA/m, a torque ($\tau = \boldsymbol{M} \times \boldsymbol{H_{AC}}$) of $4\times10^{-20}$ N$\cdot$m / $\sqrt{\textrm{Hz}}$ would correspond to a magnetic moment ($\boldsymbol{M}$) sensitivity of only $3\times10^6 \mu_B$ / $\sqrt{\textrm{Hz}}$ spins for a pure torsional mode.  Straightforward modifications to the torsional resonator geometry within this scenario pave the way to orders of magnitude improvement in torque sensitivity over the results presented here.

\begin{figure}[b]
\centerline{\includegraphics[width=3.4in]{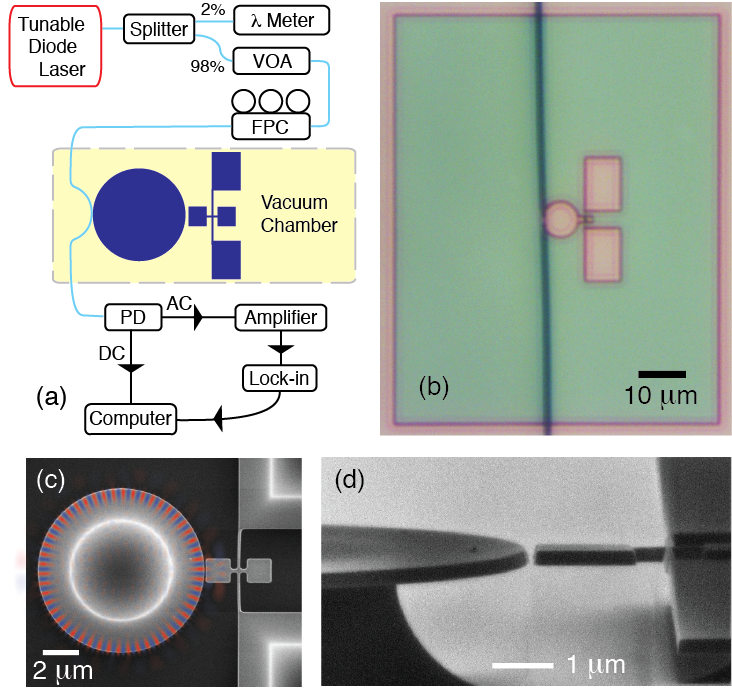}}
\caption{{\label{fig1}} (a) The schematic of the experimental setup. (b) Optical image of the dimpled, tapered fiber coupling into the (10 $\mu$m) optical microdisk of a device with a 100 nm gap to the torsion paddles. (c) Top down scanning electron microscope (SEM) image the same device as in (b) overlaid with a 2D simulation of the optical mode in the whispering gallery microdisk resonator using the effective refractive index method.  (d) SEM image of the same device tilted to 84 degrees. }
\end{figure}
 
The first step in optomechanical transduction is the formation of an optical cavity, whether it be a Fabry-Perot cavity, \cite{San10} whispering gallery mode resonator, \cite{Eic07,Par09,Sri11} Mach-Zehnder interferometer, \cite{Fon10,Sau12} or photonic crystal cavity. \cite{Eic09b,Ale11,Sun12}  Resonances occur in the optical cavity based on the geometry and effective index of refraction occupied by the optical modes.  Introduction of material with higher index than vacuum, such as a nanomechanical resonator, into the evanescent field of the optical mode shifts the effective index of refraction and hence the frequency of the resonance.  Monitoring the time dependence of the optical resonance is therefore a readout of the motion of the nanomechanical device.  This can be extremely sensitive as the optical resonances' quality factor ($Q_{opt}$) can be quite high, \cite{Hof10,Han12} the optomechanical coupling can be designed to be large,\cite{Zha11} and optical detection is free from many of the problems associated with high frequency electronic measurements. \cite{Li07b}

The layout of our optomechanical devices consists of an optical microdisk whispering gallery mode resonator side-coupled to a torsional mechanical resonator, Fig.~1c.  The torsional resonator has a torsion rod of length $\sim$5 $\mu$m and width $\sim$200 nm, and a pair of paddles that are approximately 1.5 $\mu$m on each side.  The left paddle follows the radius of curvature of the microdisk, so as to maximize coupling to the evanescent field.  Nine devices are compared for which two parameters, the optical microdisk size and the gap between the optical resonator and the mechanical resonator, are varied.  This allows experimental determination of the conditions that result in the highest optomechanical coupling and therefore lowest noise floor.  

\begin{figure}[t]
\centerline{\includegraphics[width=3.4in]{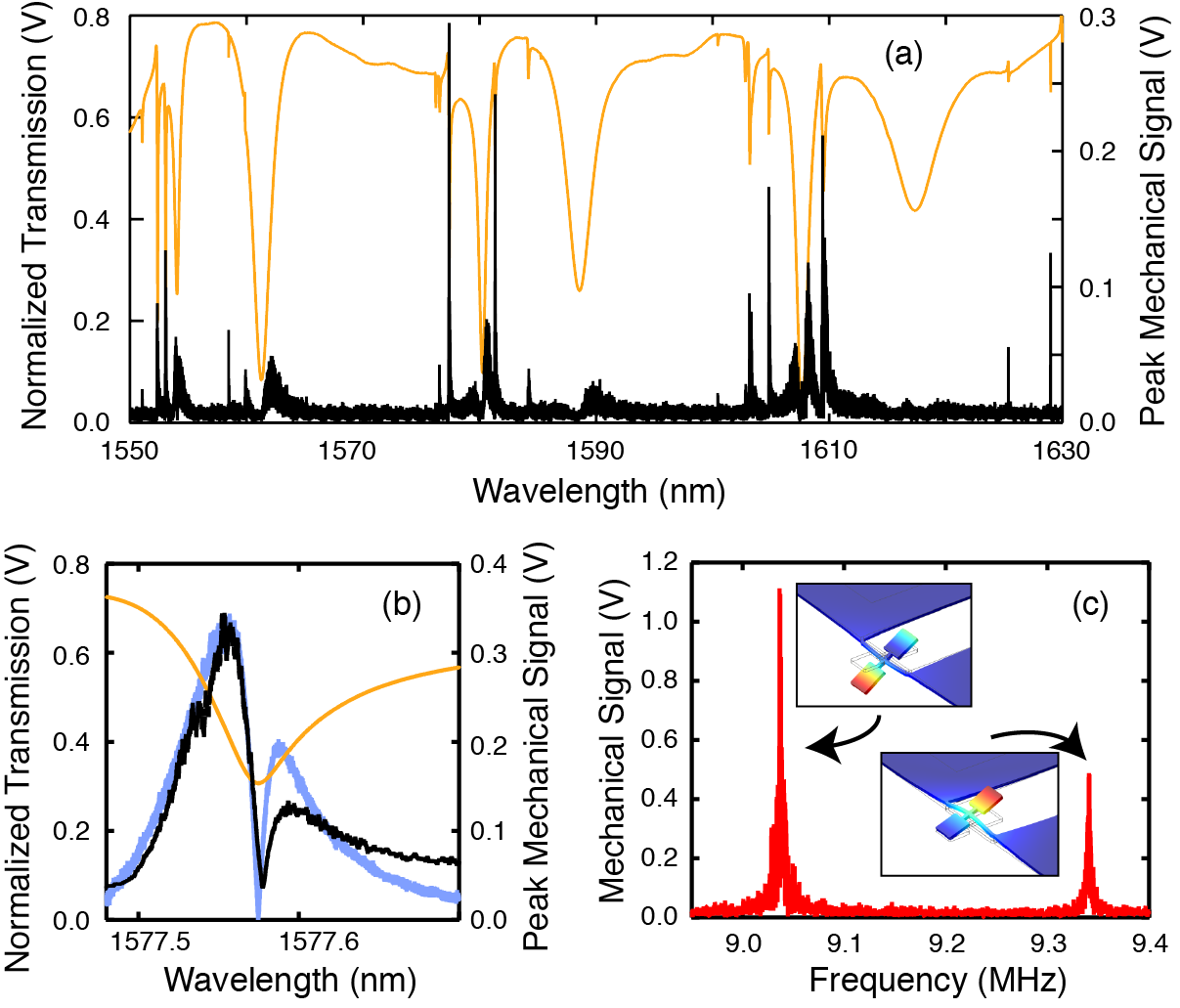}}
\caption{{\label{fig2}} (a) Coarse wavelength scan of the tunable diode laser showing the optical resonances (orange) of the device in Fig.~1, with simultaneous peak thermomechanical torsion amplitude (black) recorded using a high frequency lock-in amplifier.  (b) Same as in (a) but over a shorter wavelength range with finer resolution.  In addition we show in blue the absolute value of the scaled and smoothed numerical derivative of the optical resonance.  In both (a) and (b) the mechanical data has been smoothed - necessary as the entire wavelength sweep is performed quickly ($\sim$60 s). (c) Mechanical resonances detected optomechanically, with their corresponding mode shapes as determined using FEM (enhanced online).}
\end{figure}

The devices are fabricated from silicon-on-insulator, with a single-crystal silicon device layer of 215 nm - measured by SEM.  The device structure is patterned into the silicon by high resolution optical lithography and subsequent etching, via a commercial foundry.  The wafer is then diced into individual chips, which are released using a buffered oxide etch of the 2 $\mu$m silicon dioxide layer.  Upon subsequent drying, an in-plane distortion of the torsional resonator towards the optical microdisk is introduced, apparent in the SEM image of Fig.~1c. 

 Additionally, tilted SEM imaging reveals that the sidewalls of the both the optical and mechanical resonator are not  vertical, but instead angled by approximately 10 degrees (Fig.~1d).  This has important consequences.  For symmetric and aligned evanescent field and mechanical resonator the optomechanical coupling is second order in the displacement out of plane.  Asymmetry in the mechanical resonator \cite{Li09} and the evanescent field\cite{Gav12} independently introduce first order coupling, and account for the effectiveness of our optomechanical transduction of the torsional and out-of-plane mechanical modes.

A schematic of our fiber-based setup to measure the devices is shown in Fig.~1a.  Light from a tunable diode laser (New Focus TLB-6330) is fiber-coupled into a custom-built vacuum chamber (at $\sim$$10^{-6}$ torr) after passing through a variable optical attenuator (VOA) and fiber polarization controller (FPC).  Inside the chamber a dimpled, tapered optical fiber \cite{Eic07,Mic07} is held fixed on a goniometer and accurately aligned with the optomechanical torsional devices, which are mounted on three axes of vacuum-compatible nanopositioning stages.  The chamber has a top viewport to allow observation of the fiber while positioning the dimple to either hover above, or, to reduce mechanical instability, touch the optical whispering gallery mode resonator, Fig.~1b.  The transmission through the optical fiber is detected using a photodiode (PD: New Focus 1811) that splits the alternating (AC) and direct (DC) signals.  The DC transmission is used to monitor the amount of light coupling into the optical microdisk, whereas the AC signal, which encodes the motion of the torsional resonator, is sent to a high frequency lock-in amplifier (Zurich Instruments HF2LI).  The optical power is kept low ($\leq 20~\mu$W) to reduce optical nonlinearities. 

\begin{figure}[b]
\centerline{\includegraphics[width=3.3in]{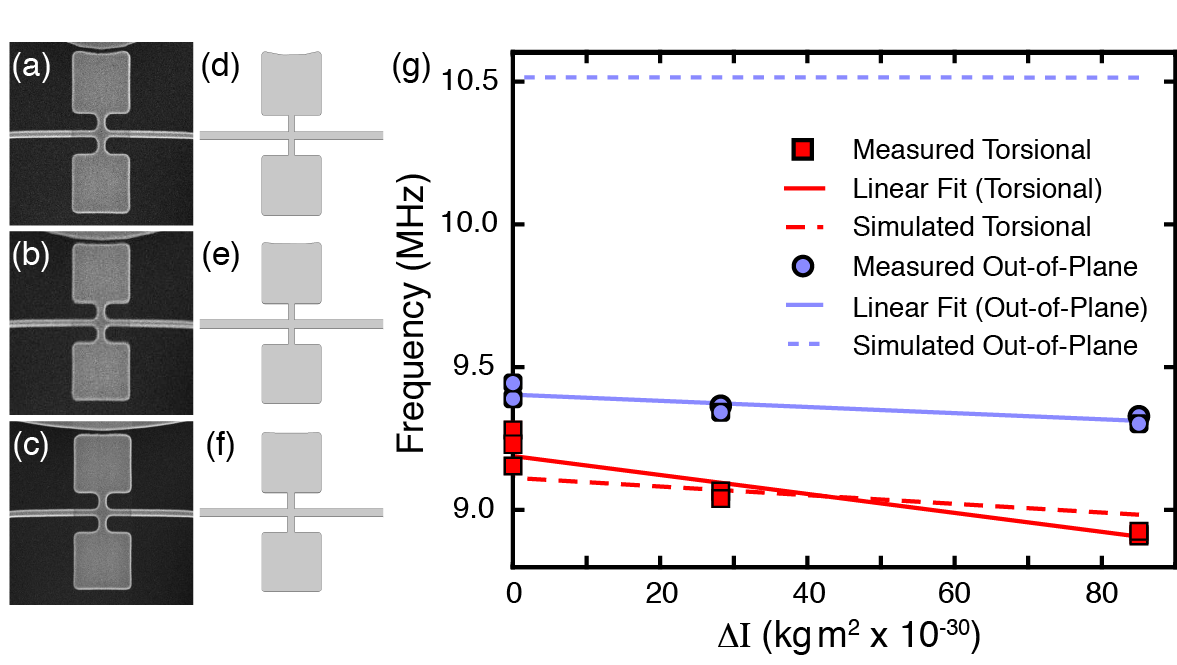}}
\caption{{\label{fig3}} SEM images of the paddles of the fabricated torsional devices with (a) 5 $\mu$m, (b) 10 $\mu$m, and (c) 20 $\mu$m optical microdisks.   (d - f)  Top-down geometries used to model the eigenfrequencies of the respective devices.  (g) Predicted frequencies calculated through FEM modeling (dashed lines), and linear fits to the measured frequencies (solid lines), of devices with varying sizes of optical microdisks plotted against the change in moment of inertia calculated from the modeled geometries (d - f). }
\end{figure}

\begin{figure*}[t]
\centerline{\includegraphics[width=\textwidth]{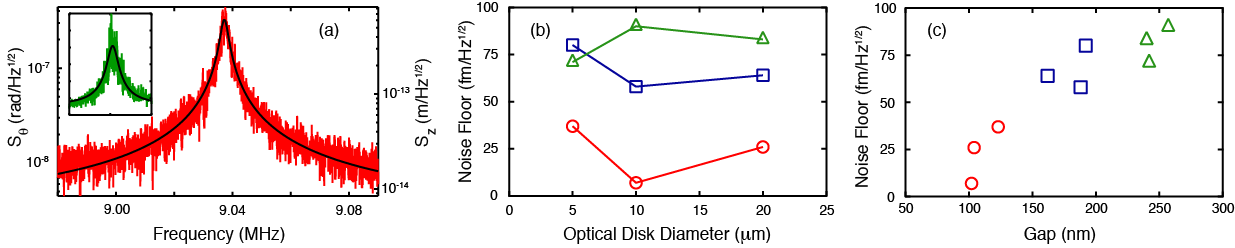}}
\caption{{\label{fig4}} (a) Calibrated thermomechanical power spectral density of a torsional resonator, with a $Q_{mech}$ of 3800, detected optomechanically in vacuum.  In the inset, the thermomechanical torsional resonance is shown at ambient pressure from 8.8 to 9.2 MHz with an arbitrary linear vertical scale.  In air the $Q_{mech}$ is 150.  The data has been smoothed by sliding average over 10 data points.  (b) Displacement noise floor versus optical microdisk diameter, with varying gap between the mechanical and optical resonator.  The red circles are for small gaps, the blue squares are the medium range gaps, and the green triangles are the large gaps.  The exact gaps can be deduced by comparison with panel (c).  (c) Displacement noise floor versus gap between the mechanical and optical resonator for all devices.  Decreasing the gap improves the optomechanical transduction efficiency - decreasing the noise floor.}
\end{figure*}

As the wavelength of the tunable diode laser is swept, optical resonances appear as dips in the DC transmission, Fig.~2a.  Simultaneously, the demodulated AC signal from the lock-in, with bandwidth of 100 Hz, is monitored to determine those optical resonances with the largest optomechanical coupling, which is dictated by the optical mode shape, the coupling between the fiber and the microdisk, and the $Q_{opt}$ of the microdisk.  This lock-in technique also allows us to directly verify the position within the optical resonance with the strongest optomechanical signal.   As can be seen in Fig.~2b, this corresponds to the largest slope of the optical resonance (which confirms the basis of the optomechanical transduction scenario described above).  A direct comparison of the peak mechanical signal with the absolute value of the numerical derivative of the optical resonance (arbitrarily scaled) shows very good qualitative agreement.

Once the wavelength of maximum optomechanical sensitivity is determined, the mechanical resonances can be further explored using the lock-in as a spectrum analyzer.  A typical frequency spectrum is shown in Fig.~2c for the device with the 10 $\mu$m diameter microdisk and 100 nm gap.  Two peaks are seen, which can be compared to finite element method (FEM) modeling of the mechanical structure using as-fabricated dimensions extracted from SEM images, Fig.~3a-c.  Eigenfrequencies are simulated for each of the three mechanical device paddle shapes, Fig.~3d-f, which are dictated by the diameter of the optical microdisk.  We have also explored how the frequencies depend on the mechanical device geometry, for example by an overall scaling factor to the lateral dimensions, or by adjusting the thickness of the silicon layer, but we find best agreement with the measured modes using the dimensions measured by SEM.  The results are shown in Fig.~3g as a function of the change in moment of inertia of the torsional resonator, $\Delta I$, calculated within the FEM model for the three paddle shapes.  The devices with a 20 $\mu$m diameter optical microdisk, as in Fig.~3c and 3f, have the smallest moment of inertia and are set as zero $\Delta I$.  In FEM simulations the lower frequency mode always corresponds to the torsional mode and the higher frequency always corresponds to the out-of-plane mode, Fig.~3g.  Yet both of these modes are substantially hybridized, in that the lower frequency torsion mode has a significant out-of-plane character and the higher frequency out-of-plane mode has significant twist, as can be seen in the simulated mode shapes in the inset to Fig.~2c.  To large part, this is due to the device undercut, which makes the effective length of the out-of plane mode longer and lowers its frequency until it is nearly degenerate with the torsional mode.  (The total out-of-plane displacement in the lower frequency mode is small, and therefore we neglect this in our calibration of the torque sensitivity.)  As the paddle shape changes, altering the moment of inertia, the frequencies of the simulated lower frequency torsion modes are noticeably shifted.  In contrast, the frequencies of the simulated higher frequency out-of-plane modes show little dependence on $\Delta I$.  This validates the identification of the measured lower frequency mode as the predominantly torsional mode, since it shows a larger dependence on $\Delta I$ than the higher frequency mode.  Further confirmation of correct mode identification comes from the fact that the higher frequency out-of-plane mode is more efficiently actuated with an external piezoelectric buzzer (that vibrates the chip out-of-plane) than the lower frequency torsional mode (not shown) - consistent with previous observations.\cite{Dav10}  Finally, we note that we observe a resonance at $\sim$$12$ MHz, which we identify as an in-plane resonance, but do not concern ourselves further with this mode here.

For all of our devices thermomechanical motion of the mechanical modes is easily resolved and external actuation is unnecessary.  Therefore calibration of the absolute displacement of the paddles is possible, enabling determination of the absolute noise floor - an important metric of transduction efficiency.  To do so a known torque is applied to the FEM simulated device, resulting in a calculated displacement and hence the torsion constant of the mode. \cite{Sad95}  Once this torsion constant is known, calibration is straightforward.  Fitting to the power spectral density gives the quality factor of the mechanical mode, $Q_{mech}$, the resonance frequency, the conversion from voltage to displacement, and the noise floor. \cite{Suh12}  A calibrated power spectral density, $S_\theta$ ($S_Z$), is shown in Fig.~4a for our best device.

Comparing the calibrated displacement noise floor across the nine different devices is useful for understanding how to optimize  torsional optomechanical coupling.  For example, shown in Fig.~4b, the best sensitivity generally occurs for 10 $\mu$m diameter optical microdisks.  Because of competing effects, this isn't obvious \emph{a priori}.  The optomechanical coupling is expected to increase as the intrinsic (unloaded) $Q_{opt}$ increases, which suggests that the largest microdisks should be best as they are least perturbed by surface effects.\cite{Bor05}  The loaded (with nanomechanical resonator present) $Q_{opt}$ of the resonances used for optomechanical detection are approximately $5\times10^3$, $1\times10^4$ and $5\times10^4$ for the 5, 10 and 20 $\mu$m microdisks respectively.   On the other hand, small microdisks have smaller optical mode volumes and therefore larger per photon optical energy densities, as well as larger overlaps between the optical field and paddle, \cite{Sri11} both of which will result in increased optomechanical coupling.  The balance between these competing effects is revealed in Fig.~4b.  However, the trend with gap between the optical microdisk and torsional resonator, Fig.~4c, is straightforward.  As the gap is decreased there is an increase in the optomechanical transduction efficiency, and decrease in the displacement noise floor, due to increasing the overlap between the mechanical resonator and the evanescent field. 

Our most sensitive device, with a 10 $\mu$m optical microdisk diameter and a gap to the torsion paddle of 100 nm, has a displacement noise floor of 7 fm/$\sqrt{\textrm{Hz}}$.  This surpasses previous experiments on macroscopic silicon torsional resonators \cite{Tit99,Hah04} and corresponds to an angular displacement noise floor 4 nrad/$\sqrt{\textrm{Hz}}$.  In a reasonable AC magnetic field strength of 1 kA/m (root mean squared), for torque actuation, this device would be sensitive to $3\times10^6 \mu_B / \sqrt{\textrm{Hz}}$ - two orders of magnitude better than recent experiments with a nanoscale torsional resonator \cite{Dav10} that had an approximately twenty times smaller torsion constant ($5\times10^{-13}$ N$\cdot$m/rad) than our current device ($1\times10^{-10}$ N$\cdot$m/rad).  Therefore simply altering our mechanical device geometry to match this torsion constant (by making the torsion rod thinner and longer for example) the $10^5\mu_B$ level could be reached.  Even further improvements to the torsional device geometry are well within the limits of current fabrication technology but have not been actively pursued up to this point, as previous measurements were detection limited and not device limited.  With the introduction of optomechanical transduction this system is now device limited.

Finally, while the $Q_{mech}$ of the devices in a vacuum of $\approx$$10^{-6}$ torr average $\sim$4000, at ambient pressure this drops to $\sim$150.  Nonetheless, thermomechanical motion of the devices can be detected in air, as shown in the inset of Fig.~4a.

Nano-optomechanical torsional resonators have been fabricated and measured.  With a displacement noise floor down to 7 fm/$\sqrt{\textrm{Hz}}$, this opens up the door for more sensitive torque measurements than have previously been possible using nanoscale torsional resonators.  In addition, straightforward optimization of the torsional resonator geometry will enable considerable improvement, based on softening of the torsion constant.  Nano-optomechanical torsional devices such as these will be useful for a number of future studies, such as exploration of nanoscale magnetic materials, or for low temperature experiments exploring quantum effects.\cite{Saf12} 

This work was supported by the University of Alberta, Faculty of Science; the Canada Foundation for Innovation; the Natural Sciences and Engineering Research Council, Canada; and the Canada School of the Energy and Environment.  Device fabrication was organized and subsidized through CMC Microsystems.  We thank Greg Popowich, Don Mullin and the University of Alberta NanoFab staff for technical assistance, Doug Vick for electron microscopy, and Marcelo Wu for helpful discussions on FEM modeling.


\begin{thebibliography}{xxx}

\bibitem{Cav98}
H. Cavendish, \textit{Philos. Trans. R. Soc. London} \textbf{88}, 469 (1798).

\bibitem{Lut82}
G.G. Luther and W.H. Towler, \textit{Phys. Rev. Lett.} \textbf{48}, 121 (1982).

\bibitem{Cle98}
A.N. Cleland and M.L. Roukes, \textit{Nature} \textbf{392}, 160 (1998).

\bibitem{Cha05}
M.D. Chabot, J.M. Moreland, L. Gao, S.H. Liou and C.W. Miller, \textit{J. Microelectromech. Systems} \textbf{14}, 1118 (2005).

\bibitem{Dav10}
J.P. Davis, D. Vick, D.C. Fortin, J.A.J. Burgess, W.K. Hiebert and M.R. Freeman, \textit{Appl. Phys. Lett.} \textbf{96}, 072513 (2010).

\bibitem{Dav10b}
J.P. Davis, D. Vick, J.A.J. Burgess, D.C. Fortin, P. Li, V. Sauer, W.K. Hiebert and M.R. Freeman, \textit{New Journal of Physics} \textbf{12}, 093033 (2010).

\bibitem{Cha01}
H.B. Chan, V.A. Aksyuk, R.N. Kleiman, D.J. Bishop and F. Capasso, \textit{Science} \textbf{291}, 1941 (2001).

\bibitem{Kle85}
R.N. Kleiman, G.K. Kaminsky, J.D. Reppy, R. Pindak and D. J. Bishop, \textit{Rev. Sci. Instrum.} \textbf{56}, 2088 (1985).

\bibitem{Bis80}
D.J. Bishop and J.D. Reppy, \textit{Phys. Rev. B} \textbf{22}, 5171 (1980).

\bibitem{Con64}
J.H. Condon and J.A. Marcus, \textit{Phys. Rev.} \textbf{134}, A446 (1964).

\bibitem{Lup99}
C. Lupien, B. Ellman, P. Gr\"utter and L. Taillefer, \textit{Appl. Phys. Lett} \textbf{74}, 451 (1999).

\bibitem{Li07}
L. Li, J. G. Checkelsky, S. Komiya, Y. Ando and N. P. Ong, \textit{Nat. Phys.} \textbf{3}, 311 (2007).

\bibitem{Ble09}
A.C. Bleszynski-Jayich, W.E. Shanks, B. Peaudecerf, E. Ginossar, F. von Oppen, L. Glazman and J.G.E. Harris, \textit{Science} \textbf{326}, 272 (2009).

\bibitem{Bet36}
R. Beth, \textit{Phys. Rev.} \textbf{50}, 115 (1936).

\bibitem{Eki05}
K.L. Ekinci and M.L. Roukes, \textit{Rev. of Sci. Inst.} \textbf{76} (2005).

\bibitem{Ane09}
G. Anetsberger, O. Arcizet, Q.P. Unterreithmeier, R. Rivi\'ere, A. Schliesser, E.M. Weig, J.P. Kotthaus and T. J. Kippenberg, \textit{Nat. Phys.} \textbf{5}, 909 (2009).

\bibitem{Eic09}
M. Eichenfield, R.M. Camacho, J. Chan, K.J. Vahala and O. Painter, \textit{Nature} \textbf{459}, 550 (2009).

\bibitem{Ane10}
G. Anetsberger, E. Gavartin, O. Arcizet, Q.P. Unterreithmeier, E.M. Weig, M.L. Gorodetsky, J.P. Kotthaus and T.J. Kippenberg, \textit{Phys. Rev. A} \textbf{82}, 061804R (2010).

\bibitem{Bry03}
Z. Bryant, M.D. Stone, J. Gore, S.B. Smith, N.R. Cozzarelli and C. Bustamante, \textit{Nature} \textbf{424}, 338 (2003).

\bibitem{Lip10}
J. Lipfert, J.W.J.J. Kerssemakers, T. Jager and N.H. Dekker, \textit{Nat. Methods} \textbf{7}, 977 (2010).

\bibitem{San10}
J.C. Sankey, C. Yang, B.M. Zwickl, A.M. Jayich and J.G.E. Harris, \textit{Nat. Phys.} \textbf{6}, 707 (2010).

\bibitem{Eic07}
M. Eichenfield, C.P. Michael, R. Perahia and O. Painter, \textit{Nat. Photonics} \textbf{1}, 416 (2007).

\bibitem{Par09}
Y.-S. Park and H. Wang, \textit{Nat. Phys.} \textbf{5}, 489 (2009).

\bibitem{Sri11}
K. Srinivasan, H. Miao, M.T. Rakher, M. Davan\c{c}o and V. Aksyuk,  \textit{Nano. Lett} \textbf{11}, 791 (2011).

\bibitem{Fon10}
K.Y. Fong, W.H.P. Pernice, M. Li and H.X. Tang, \textit{Appl. Phys. Lett} \textbf{97}, 073112 (2010).

\bibitem{Sau12}
V.T.K. Sauer, Z. Diao, M.R. Freeman and W.K. Hiebert, \textit{Appl. Phys. Lett} \textbf{100}, 261102 (2012).

\bibitem{Eic09b}
M. Eichenfield, J. Chan R.M. Camacho, K.J. Vahala and O. Painter, \textit{Nature} \textbf{462}, 78 (2009).

\bibitem{Ale11}
T.P.M. Alegre, A. Safavi-Naeini, M. Winger and O. Painter, \textit{Optics Express} \textbf{19}, 5658 (2011).

\bibitem{Sun12}
X. Sun, J. Zheng, M. Poot, C.W. Wong and H.X. Tang, \textit{Nano Lett.} \textbf{12,}, 2299 (2012).

\bibitem{Hof10}
J. Hofer, A. Schliesser and T.J. Kippenberg, \textit{Phys. Rev. A} \textbf{82}, 031804R (2010).

\bibitem{Han12}
H. Lee, T. Chen, J. Li, K.Y. Yang, S. Jeon, O. Painter and K.J. Vahala, \textit{Nat. Photonics} \textbf{6}, 369 (2012).

\bibitem{Zha11}
M. Zhang, G. Wiederhecker, S. Manipatruni, A. Barnard, P. McEuen and M. Lipson, arXiv:1112.3636v1 (2011).

\bibitem{Li07b}
M. Li, H.X. Tang and M.L. Roukes, \textit{Nat. Nano.} \textbf{2}, 114 (2007).

\bibitem{Gav12}
E. Gavartin, P. Verlot and T.J. Kippenberg, \textit{Nat. Nano.} \textbf{7}, 509 (2012).

\bibitem{Li09}
M. Li, W.H.P. Pernice and H.X. Tang, \textit{Nat. Nano.} \textbf{4}, 377 (2009).

\bibitem{Mic07}
C.P. Michael, M. Borselli, T. J. Johnson, C. Chrystal and O. Painter, \textit{Optics Express} \textbf{15}, 4745 (2007).

\bibitem{Sad95}
J.E. Sader, I. Larson, P. Mulvaney and L.R.White, \textit{Rev. Sci. Instrum.} \textbf{66}, 3789 (1995).

\bibitem{Suh12}
A. Suhel, B.D. Hauer, T.S. Biswas, K.S.D. Beach and J.P. Davis, \textit{Appl. Phys. Lett.} \textbf{100}, 173111 (2012).

\bibitem{Bor05}
M. Borselli, T.J. Johnson and O. Painter, \textit{Optics Express} \textbf{13}, 1515 (2005).

\bibitem{Tit99}
I. Tittonen, G. Breitenbach, T. Kalkbrenner, T. M\"uller, R. Conradt, S. Schiller, E. Steinsland, N. Blanc and N. F. de Rooij, \textit{Phys. Rev. A} \textbf{59}, 1038 (1999).

\bibitem{Hah04}
O. Hahtela, K. Nera and I. Tittonen, \textit{Jour. of Opt. A} \textbf{6}, S115 (2004).

\bibitem{Saf12}
A.H. Safavi-Naeini, J. Chan, J.T. Hill, T.P. Mayer Alegre, A. Krause and O. Painter, \textit{Phys. Rev. Lett.} \textbf{108}, 033602 (2012).


\end{thebibliography}
\end{document}